\documentclass[11pt,a4]{article}          

\usepackage[T1]{fontenc}

\usepackage{graphicx}
\usepackage{mathptmx}      
\usepackage{flushend}
\usepackage{wasysym}
\usepackage[numbers,sort&compress]{natbib}
\usepackage[colorlinks,citecolor=blue,urlcolor=blue,linkcolor=blue]{hyperref}

\usepackage{lineno}

\begin{document}
\renewcommand{\dagger}{\circ}
\title{Feasibility Studies of the Diffractive Bremsstrahlung Measurement at the LHC
\author{
J. J. Chwastowski$^a$\footnote{Corresponding author, e-mail: Janusz.Chwastowski@ifj.edu.pl}\ , S. Czekierda$^a$,  R. Kycia$^b$,\\
R. Staszewski$^a$, J. Turnau\footnote{\textit{Professor emeritus}}\ , M. Trzebinski$^a$\\[8pt]
$^a$The Henryk Niewodnicza\'nski Institute of Nuclear Physics,\\
Radzikowskiego 152,
31-342 Krak\'ow, Poland\\[8pt]
$^b$Dept. of Physics, Mathematics and Computer Science,\\
Cracow University of Technology,
Warszawska 24,\\
31-155 Krak\'ow, Poland\\[8pt]
}
}


\maketitle

\begin{abstract}
Feasibility studies of an observation of the exclusive diffractive bremsstrahlung in proton-proton scattering at the LHC are reported. 
A simplified approach to the photon and the scattered proton energy reconstruction is used. The background influence is discussed. 
\end{abstract}

\section{Introduction}
Electromagnetic bremsstrahlung is widely used in various applications. They range from medicine to advanced methods in structural research.
In high energy physics the bremsstrahlung photons were used, \textit{inter alia}, to study photoproduction -- see for example \cite{photoprod}. With 
advent of HERA, it became a very attractive tool. On one hand the bremsstrahlung events are characterised by a simple and easy to register final state, 
and on the other hand, in the HERA energy range, the process cross-section is relatively large and precisely calculable. Therefore, bremsstrahlung was 
used to determine the machine absolute and instantaneous luminosities and in addition it served as an efficient beam diagnostic and monitoring tool  (see  
\cite{zeus}). The measurements exploited the basic properties of the final state: an approximate energy conservation as well as the angular properties of 
radiated photons.

In 2011, V. Khoze \textit{et al.} \cite{khoze1} postulated an extension of the LHC physics programme with the observation of the high energy photon 
radiated 
in the elastic proton-proton scattering event -- the study of exclusive diffractive bremsstrahlung. The measurements of exclusive diffractive bremsstrahlung 
can be considered as complementary to the luminometers and luminosity monitors proposed in \cite{khoze2, krasny1}. It is also worth noting that this 
process has never been studied at high energies. The difference with respect to the electromagnetic bremsstrahlung is that the interaction is mediated by 
an exchange of a pomeron. In both electromagnetic and diffractive  processes the exchange is colourless and soft, and the sum of the radiated photon 
and the scattered proton energies is nearly equal to the beam energy. The calculations of \cite{khoze1} were  considerably extended by P. Lebiedowicz and 
A. Szczurek \cite{szczurek} who also introduced the proton formfactor into the calculations. However, the values of the parameters in their model are 
based on an educated guess and are subject to experimental verification. The authors of \cite{szczurek} used the Donnachie-Landshoff parameterisation 
\cite{dl} with linear pomeron trajectory with the intercept 1.0808 and the slope $\alpha^\prime_{I\!P} = 0.25$ GeV$^{-2}$. The elastic scattering slope 
evolution is given by $B(s) = B^{NN}_{I\!P}+2\cdot\alpha^\prime_{I\!P}\ln{s/s_0}$ where $s_0 = 1$ GeV$^2$ and $B^{NN}_{I\!P} = 9$ GeV$^{-2}$.

P. Lebiedowicz and A. Szczurek considered also other mechanisms leading to the $p\,p\,\gamma$ final state, for example the virtual photon re-scattering. 
They concluded that these processes do not play a very important role and can be safely neglected when considering the forward directions. For 
a full account of the model details and its parameters an interested reader is referred to original publications. 

In this paper a feasibility study of the diffractive photon bremsstrahlung measurement in proton-proton interactions at the presently available centre of 
mass energy of 13 TeV at the LHC is reported. Registration of bremsstrahlung photons in the ATLAS \cite{atlas} very forward Zero Degree Calorimeter 
(ZDC) \cite{zdc} and the scattered protons in the ATLAS Forward Proton detector \cite{afp} is considered. 

\section{Experimental Set-up}

\subsection{Accelerator}

During the Long Shutdown 1  the LHC underwent several improvements. In consequence, it delivers proton beams accelerated to 6500 GeV with high 
instantaneous luminosity. 

The operation mode of the machine can be defined by a set of parameters defining the properties of its magnetic lattice -- the so-called accelerator optics. 
At the LHC the name of the set comes after the value of the betatron function\footnote{The betatron function defines the distance measured from a given 
point along the orbit, after which the beam dimensions in the transverse plane to the motion are doubled.} at the Interaction Point (IP), $\beta^\star$. 
The nominal or ``collision'' optics is characterised by the  $\beta^\star$ value of 0.55~m. Table \ref{tab:param} presents the selected parameters of the 
proton beams for the beam emittance\footnote{The beam emittance measures the spread of the beam particles in the momentum-position phase space.}, 
$\epsilon = 3.75$ $\mu$rad\,m, collision optics and the beam energy $E_{beam} = 6500$ GeV.

\begin{table}[h]
\centering
\caption{The beam parameters: the  angular dispersion, the beam transverse size at the detector location and the crossing angle for the collision optics 
for the beam energy $E_{beam} = 6500$ GeV.} 
\label{tab:param}

\begin{tabular*}{\columnwidth}{@{\extracolsep{\fill}}ccc@{}}
 {Angular}  &	{Beam transverse}&Beam crossing\\
 {dispersion [$\mu$rad] } & {size [mm]}  &  half angle [$\mu$rad]\\
\hline
31.4& 0.19 & -142.5\\
\hline
\end{tabular*}
\end{table}

\begin{figure*}
\begin{minipage}{\columnwidth}
\centering
\includegraphics[width=\columnwidth]{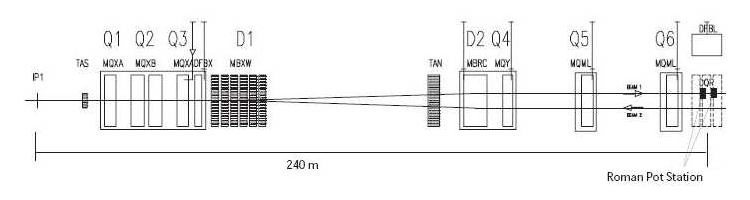}
\end{minipage}
\caption{The LHC magnetic lattice.}
\label{fig:lhc}
\end{figure*}

The LHC accelerates two proton beams. The one performing clockwise rotation is called  \textit{beam1}, and the other one -- \textit{beam2}. The beams 
traverse the LHC magnetic lattice in separate beam pipes which merge into a single one about 140 m away from the Interaction Point (IP). A schematic 
view of the LHC lattice in the vicinity of the ATLAS IP is shown in Fig.~\ref{fig:lhc}.
The quadrupole magnets are labelled with ``Q'', and the dipole ones -- with ``D''. The Q1--Q3 quadrupoles serve as the final focusing and the emittance 
matching triplet, while the D1 and D2 dipoles are used to separate the proton beams. In the analysis presented below the LHC magnetic lattice plays a role 
of the magnetic spectrometer for the scattered forward particles. In addition there are several corrector magnets, beam position monitors and collimators,
not shown in the figure. To protect the magnets against the radiation coming from the Interaction Point the Target Absorber Secondaries (TAS) and the 
Target Absorber Neutral (TAN) were installed in front of the final focusing triplet and the D2 magnet, respectively.

\subsection{Zero Degree Calorimeters}

Zero Degree Calorimeters (ZDC) \cite{zdc} are compact detectors foreseen to measure neutral particles produced at large pseudorapidities. They are 
located on both sides of the ATLAS IP at the distances of 140 m in the slit of the TAN absorber. Each ZDC is segmented longitudinally with the first segment 
dedicated to the measurement of electromagnetic radiation. The energy resolution of the calorimeter for photons is $\sigma/E = 0.58/\sqrt{E}+0.02$ and 
for neutrons $\sigma/E = 2.1/\sqrt{E}+0.12$ with the particle energy measured in GeV. The calorimeter is also the particle impact position sensitive.
Throughout this work, it was assumed that the ZDC spatial resolution is 1 mm for photons and 3 mm for neutrons, irrespectively of the incident particle 
energy.  These values are upper bounds on the resolutions and were estimated during the ZDC tests -- see \cite{zdc} for details. 
The active area of the ZDC face as seen from the ATLAS experiment IP  is  of an approximately rectangular shape with dimensions 94 mm by 88 mm in 
horizontal and vertical directions, respectively.

\subsection{Forward Tracker -- AFP}

Protons scattered or emitted at very low polar angles escape from the main ATLAS detector through the accelerator beam pipe and traverse the magnetic 
lattice of the machine. They can be registered in the dedicated detectors. Typically, such detectors use the roman pot technology which allows the precise 
positioning of the detectors inside the beam pipe. It is quite clear that the acceptance of the detectors depends on both the properties of the magnetic 
spectrometer and the position of the detector in the beam pipe. The latter is quantified by a crucial parameter -- the distance between the detector active 
edge and the beam. One should also note that in principle a given event can have a single or double tag signature denoting one or two scattered forward 
protons. Important variables describing the scattered proton are: the proton transverse momentum, $p_T$, and the relative energy loss, $\xi = (E_{beam} 
- E^\prime_p)/E_{beam}$, where $E^\prime_p$ is the scattered proton energy and $E_{beam}$ is the beam energy.

The ATLAS Forward Proton (AFP) detectors \cite{afp} are foreseen to work during the usual running of the LHC \textit{i.e.} in runs with 
$\beta^\star = 0.55$ m. Four AFP stations are planned. They will be placed symmetrically w.r.t. the ATLAS IP at the distances of 205 m and 217 m. Each 
station contains a silicon tracker, which can be inserted horizontally into the beam pipe. The tracker contains four planes of silicon pixel detectors. The 
planes are tilted w.r.t. the $x$ axis (horizontal direction) and staggered in the $y$ axis (vertical) direction. The resulting spatial resolution of the scattered 
proton 
track measurement is about 10 $\mu$m and 30 $\mu$m in the horizontal and vertical direction, respectively. The distance between the edge of the active 
part of the detector and the outer wall (beam side) of the roman pot was assumed to be 0.5 mm. The active detector area projected along the beam 
direction is $16.3 \times 20$ mm$^2$. The energy of the scattered proton registered by the AFP detector can be determined with the resolution better 
than 10~GeV \cite{rafal}. The outer stations will contain also the time-of-flight counters providing the timing resolution of about 10 ps, and 
hence they allow the determination of the interaction vertex location with precision of about 3 mm. Timing detectors will be of primordial importance in 
high instantaneous luminosity runs when high pile-up will be present; however, they do not play any role in the discussed analysis.

In the case of the collision optics, the AFP detector acceptance is large for the scattered protons having the relative energy loss within the range 
$(0.02, 0.12)$ \cite{maciek}.

\section{Properties of Final State}

The exchanged pomeron ensures the four-momentum conservation in the diffractive bremsstrahlung process:
$$
p + p \rightarrow p+p+\gamma,
$$
and since it is soft, the equation
\begin{equation}
E_{beam} \approx E_\gamma+E^\prime_p,
\label{eq:econs}
\end{equation}
 where $E^\prime_p$ is the scattered proton energy, is fulfilled with a very good accuracy.
The process cross-section is of the order of a few microbarns. Moreover, similarly to the electromagnetic bremsstrahlung, the photon angular distribution 
is driven by the radiating particle Lorentz factor $\gamma = E_p/m_p$ 
 $$ \frac{d\sigma}{d\Theta_\gamma} \sim \frac{\Theta_\gamma} {\left(\frac{m_p^2}{E_p^2}+\Theta_\gamma^2\right)^2},$$
 \noindent
where $\Theta_\gamma$ is the polar angle of the emitted photon, $m_p$ is the proton mass and $E_p = E_{beam}$ its energy \cite{jackson}. 

The average value of the photon polar angle is about $1/\gamma$. In the case of the radiating proton accelerated to the energy of 6500 GeV, this angle is 
of the order of 144 $\mu$rad, which corresponds to the photon pseudorapidity\footnote{The pseudorapidity of a particle is defined as 
$\eta = -\ln{\tan{\Theta/2}}$, where $\Theta$ is the polar angle of a particle.}, $\eta \approx 9.5$. One should also note that the scattered proton 
angular distribution is also extremely narrow due to the large value of the nuclear slope parameter at high energies and its average value is 
$\sim\!1/4\gamma$  at the LHC energies.

\begin{figure}[h]
\begin{minipage}{\columnwidth}
\centering
\includegraphics[width=0.8\columnwidth]{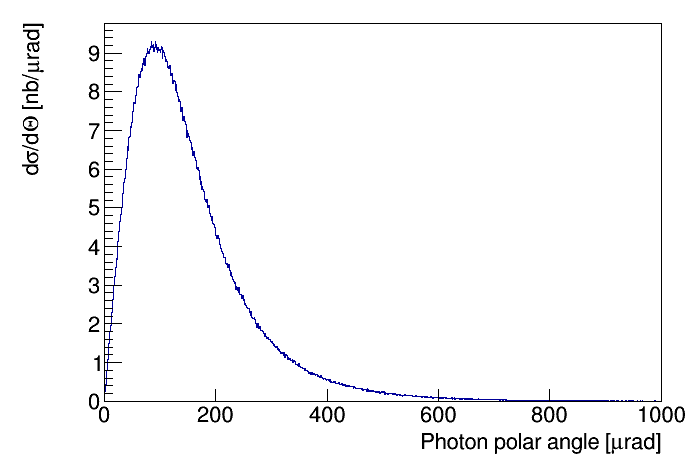}
\end{minipage}
\caption{The bremsstrahlung photon polar angle distribution.}
\label{fig:gtheta}
\end{figure}
\begin{figure}[h]
\begin{minipage}{\columnwidth}
\centering
\includegraphics[width=0.8\columnwidth]{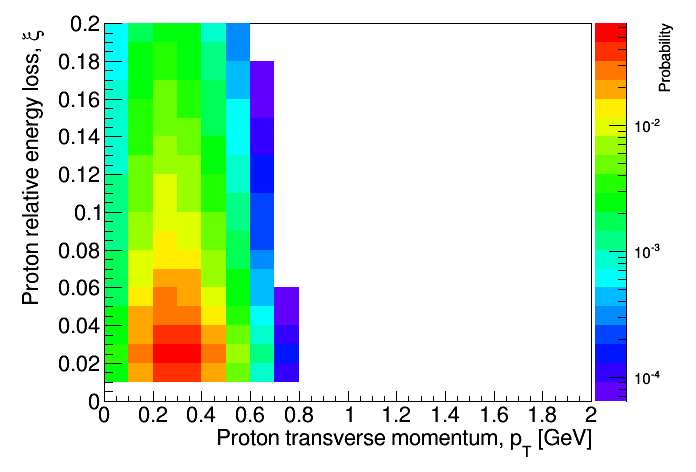}
\end{minipage}
\caption{The distribution of the radiating proton relative energy loss, $\xi$, versus its transverse momentum, $p_T$.}
\label{fig:gksi}
\end{figure}

Figure \ref{fig:gtheta} shows the distribution of the polar angle of the bremsstrahlung photon. In Fig. \ref{fig:gksi} the distribution of the radiating proton 
relative energy loss, $\xi$, versus its transverse momentum, $p_T$ is shown. The photon energy spectrum (not shown) is of a typical shape 
($\sim\!1/E_\gamma$) and that of the proton relative energy loss is proportional to $1/\xi$. The final state particles are emitted in the forward directions. 
For the LHC running with $\beta^\star = 0.55$ m it is impossible to register the elastically scattered protons with the help of the AFP stations. Therefore,
the signal signature is defined as the two-particle final state with a photon registered in the ZDC and the scattered proton in the AFP located in 
the photon hemisphere. Moreover, their reconstructed energies have to fulfill relation (\ref{eq:econs}). The energy range of the scattered proton, and 
hence also of the photon, is defined by the machine optics and location of the AFP detectors.

\section{Analysis}

\subsection{Signal}
\label{sec:signal}

A sample of the signal events was generated using the Monte Carlo generator based on the GenEx MC \cite{kycia1} which was devised to describe the 
production of the low multiplicity exclusive states. The GenEx extension implements the calculations of \cite{szczurek}. A sample of 3 000 000 events with 
the photon energy between 100 GeV and 1500 GeV was generated. The calculated cross-section within this range is $1.7514\pm0.0006~\mu$b.

The final state particles were transported through the LHC lattice. The transport of the protons was performed using the parameterisation \cite{rafal} of 
the MAD-X~\cite{mad} results. The differences between the MAD-X and the parameterisation results are much smaller than the spatial resolutions of the 
experimental apparatus. In the calculations the beam and interaction vertex properties were taken into account. One should note that the beam chamber 
geometry is of negligible importance owing to the low values of the transverse momenta of the final state particles. The ZDC response to the incident 
photon was calculated using the resolution mentioned above. Also, the photon impact position was smeared according to the reported spatial resolution of 
the device. The energy of the proton reaching the AFP was smeared with the energy unfolding resolution. 

The event selection depends on the machine optics and on the apparatus configuration.
Speaking in general terms, the signal event is defined by the following requirements:
\begin{itemize}
\item presence of a photon with the energy reconstructed in the ZDC, $E_{\gamma,ZDC}$,
\item presence, in the photon hemisphere, of a proton with the energy $E^\prime_{AFP}$,
\item lack of the proton signal in the other hemisphere.
\end{itemize} 

\subsection{Simulation of Background Processes}

In the background influence study events producing the signal-like signature were considered. This means that the central detector should not register any 
particle and there should be an electromagnetic energy deposit in the ZDC and a proton registered in the AFP located in the photon hemisphere. 
Therefore, an event was rejected from the analysis if it satisfied at least one of the following requirements:
\begin{itemize}
\item presence of a charged particle with the transverse momentum $p_T > 1$ GeV and $|\eta| < 2.5$ -- the inner tracker veto;
\item presence of a particle with energy $E > 1$ GeV and $|\eta| < 4.8$ -- the calorimeter veto,
\item presence of a neutral hadron with the energy reconstructed in any ZDC greater than 30 GeV,
\item the electromagnetic energy reconstructed in the ZDC belonging to the non-signal hemisphere larger than 30 GeV.
\end{itemize}
\noindent
The first two selection cuts efficiently reject non-diffractive and high-mass diffractive processes, while the other two play important an role in the reduction 
of the double diffraction contribution.  

The background processes were generated using \textsc{Pythia 8} Monte Carlo \cite{pythia}. The generated sample includes single and double diffractive 
dissociation\footnote{It was checked with Pythia that the non-diffractive processes contribution to the background after the final selection is less than 
0.5\permil{} of the diffractive one.}. The \textsc{Pythia} reported cross-section is 21.4 mb. A sample of $10^9$ events was generated. These events 
underwent the simulation 
and reconstruction procedures as described in Sec. \ref{sec:signal}. It was checked with \textsc{Pythia 8} that the dominating source of the 
background events is the process:
$$p+p\rightarrow p+ \pi^0 +p.$$
The events were checked for the presence of a \textit{single} photon within the ZDC acceptance. For each photon within the ZDC acceptance its 
reconstructed position was calculated assuming the ZDC spatial resolution. The events containing two or more photons were removed from the analysis if 
the maximum distance between their impact positions was larger than 6 mm. One should note that the distance between the photons emerging from the 
$\pi^0$ decay at the ZDC face is not smaller than 5 mm in the considered energy range and which is a consequence of the $\pi^0$ mass.

\subsection{Optimisation of the Signal/Background Ratio}

To simulate the measurement the following assumptions were used:
\begin{itemize}
\item energy of a photon reconstructed in the ZDC, $E_{\gamma,ZDC}$, within the range 130 GeV to 1500 GeV,
\item in the photon hemisphere a proton with the energy $E^\prime_{AFP}$ corresponding to $\xi > 0.02$  at $\sqrt{s} = 13$ TeV.
\end{itemize} 

\begin{figure}[h]
\begin{minipage}{\columnwidth}
\centering
\includegraphics[width=\columnwidth]{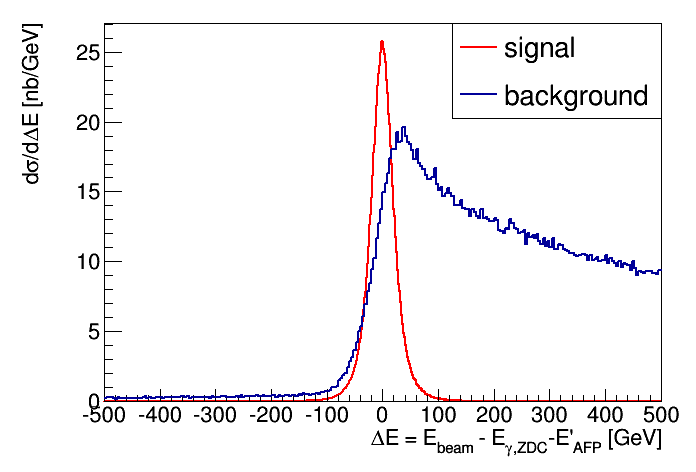}
\end{minipage}
\caption{The distribution of $\Delta E = E_{beam}-E_{\gamma,ZDC}-E^\prime_{AFP}$. The red line - the signal event sample, the blue line - the background sample (colour on-line).}
\label{fig:econs}
\end{figure}

For both signal and background samples, the distributions of $\Delta E = E_{beam}-E_{\gamma,ZDC}-E^\prime_{AFP}$, 
were constructed. They are shown in Fig. \ref{fig:econs}.

The distributions have very different shapes. The signal one is peaked at the beam energy and its width, $\sigma_E = 26$ GeV. The distribution 
constructed using the background events has a non-gaussian shape with long asymmetric tails. Also. its maximum is shifted from the beam energy value. 
Therefore the requirement of the energy conservation:
\begin{equation}
|E_{beam} - E_{\gamma,ZDC}-E^\prime_{AFP}| < 3\cdot\sigma_E,
\label{eq:expecons}
\end{equation}
efficiently rejects the background.

The above condition leads to the following visible cross-sections: $\sigma_{sig,vis} \approx 1.31$ $\mu$b for the signal and $ \sigma_{bckg,vis} \approx 
1.88~\mu$b for the background. In the following, further improvement of the signal to background ratio is attempted. This was performed using features 
of the signal and background final state particles. At the first step the differences in the distributions of the photon polar angle were studied. To quantify 
them the distributions of the photon impact position on the ZDC face were investigated for the signal and background samples.

\begin{figure}[h]
\begin{minipage}{\columnwidth}
\centering
\includegraphics[width=\columnwidth]{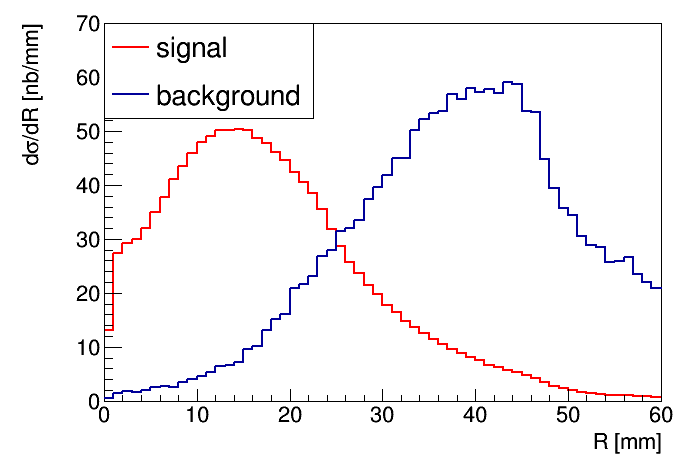}
\end{minipage}
\caption{The distribution of the distance, $R$, between the reconstructed photon impact point at the ZDC and the beam projection. The red line - the signal event sample, the blue line - the background sample (colour on-line).}
\label{fig:rsig}
\end{figure}

In particular, the distributions of the distance, $R$, measured at the ZDC face, between the photon impact point and the linearly extrapolated position of 
the beam following from the beam crossing angle -- c.f. Table \ref{tab:param} were studied. These distributions are shown in Fig. \ref{fig:rsig} for events 
passing the above outlined selections.
\noindent
The signal distribution has a clear maximum located at about 14 mm and a quickly falling tail. The distribution constructed for the background events 
increases with increasing value of the distance, reaches an approximate plateau for the distance of about 34 mm and for values larger than about 46 mm 
it quickly decreases. This difference will serve as an efficient background discrimination tool.

\begin{figure}[h]
\begin{minipage}{\columnwidth}
\centering
\includegraphics[width=\columnwidth]{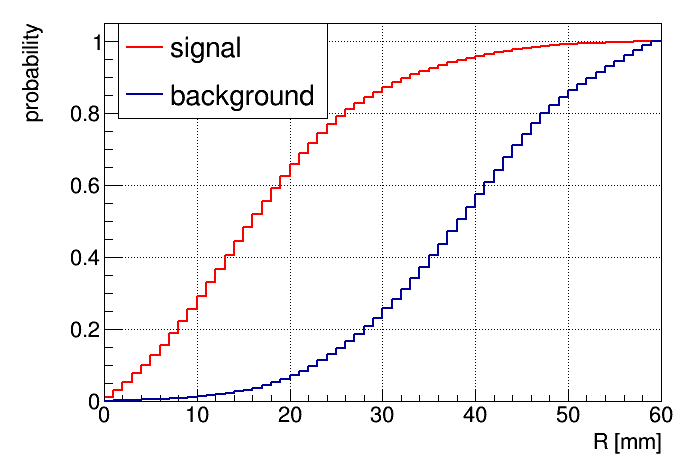}
\end{minipage}
\caption{The cumulative distribution of the probability that the distance, $R$, between the reconstructed photon impact point at the ZDC and the beam 
projection  is smaller than a certain value. The red line - the signal event sample, the blue line - the background sample (colour on-line).}
\label{fig:risig}
\end{figure}

Both distributions were converted to the distributions of the probability that the discussed distance is smaller than a certain value and  are shown in  Fig. 
\ref{fig:risig}. 

Inspection of  Fig. \ref{fig:risig} shows that the requirement that the distance being smaller than 30 mm leaves about 85\% percent of the accepted signal 
events and rejects about 75\% of the background events. Imposing such a cut leads to the final visible cross-sections of $1.12$~$\mu$b and $394$ nb for 
the signal and the background, respectively.

The above mentioned cross-section values are subject to a change due to the geometry and the position of the AFP detector with respect to the proton 
beam. As was already, mentioned the AFP acceptance crucially depends on the distance between the beam and the active edge of the detector. Typically
this distance is set to the multiple of the transverse size of the beam, $\sigma$, at the detector location (see Table \ref{tab:param}). In the calculations 
this distance was set to $10\sigma, 15\sigma, 20\sigma$ and $25\sigma$. Table \ref{tab:visxs} lists the visible cross-sections as a function of the 
mentioned distances for the signal and background samples. Also, the signal to background ratio is given in the table.

\begin{table}[h]
\centering
\caption{The visible cross-sections in nb for the signal and background for different distances of the detector active part and the beam. }
\label{tab:visxs}

\begin{tabular*}{\columnwidth}{@{\extracolsep{\fill}}lrrrrcc@{}}
 &	{Signal}  & {Background}&{S/B ratio}\\
distance& {cross-section [nb] } & {cross-section [nb]}  &  \\
\hline
$10\sigma$ &1047& 280&  3.5 \\
$15\sigma$ & 915& 291 & 3.1\\
$20\sigma$ & 745& 299 & 2.5\\
$25\sigma$ & 614& 298 & 2.1\\
$30\sigma$ & 497& 290 & 1.8 \\
\hline
\end{tabular*}
\end{table}

The geometrical acceptance of the AFP system reduces the visible cross-section to about 1 $\mu$b in the case of the closest approach of the detector to 
the beam. The visible cross-section for the signal decreases with increasing distance between the active detector edge and the beam while that for the 
background is approximately independent the detector position within the considered range of distances. These trends result in the decreasing value of the 
signal to background ratio from about 3 for the distance of 10$\sigma$ to about 2 for 30$\sigma$

\subsection{Results}

The signal visible cross-section predicts that the rate of the diffractive bremsstrahlung events within the considered phase space varies between 0.1 Hz to 
about 100 Hz for the instantaneous luminosity increasing from $10^{29}$ cm$^{-2}$s$^{-1}$ to $10^{32}$ cm$^{-2}$s$^{-1}$ assuming that the 
luminosity is evenly distributed among all bunches. Moreover, the above results were obtained with an implicit assumption of a single interaction per bunch 
crossing. In reality, at the LHC running with large instantaneous luminosity, several $pp$ interactions per bunch crossing can happen. Therefore, the visible 
cross-sections have to be multiplied by the probability of no $pp$ interactions to calculate the pile-up-free rate. The reduction factor depends on the 
machine delivered luminosity and is negligible at luminosities below $10^{30}$ cm$^{-2}$s$^{-1}$, reaches the value of about 0.9 for luminosity of 
$10^{31}$ cm$^{-2}$s$^{-1}$, the value of about 0.4 for $10^{32}$ cm$^{-2}$s$^{-1}$ and for larger instantaneous luminosities rapidly decreases.\\
\noindent
The visible cross-section implies the collection of the diffractive bremsstrahlung samples of a few thousand to several thousand events in a 10-hour long 
LHC run.

It should be noted that a similar study was performed for $pp$ scattering at RHIC energies \cite{app} using the STAR apparatus. Yet, another study was 
performed also at $\sqrt{s} = 10$ TeV and for the $\beta^\star = 90$ m \cite{khoze1} for the TOTEM. In contrast to the work presented above the double 
tagged events were considered. The authors concluded that the photons can be used to tag the elastic events which in turn can be used to measure the 
$p_T$ dependence of the elastic cross-section as well as the ratio of the elastic to total cross-section.
\section{Summary and Conclusions}

Feasibility studies of the diffractive bremsstrahlung measurement at the centre of mass energy of 13 TeV at the LHC were performed. The expected 
visible cross-section depends on the distance between the detector and the beam. It decreases with increasing distance from about 1 $\mu$b (separation
of 10$\sigma$ of the beam) to about 500 nb for 30$\sigma$. The visible cross-section for the background is approximately constant. The expected rates 
allow the collection of large event samples during a 10-hour long LHC run with moderate instantaneous luminosity.

Further optimisation of the analysis cuts has to take into account actual running conditions during the data taking period.

\section{Acknowledgements}
\ \\
We are very much indebted to P. Lebiedowicz and A. Szczurek for many stimulating discussions. This work was supported in part by Polish National Science 
Centre grant UMO-2012/05/B/ST2/02480.

\end{document}